\documentclass[12pt]{iopart}

\usepackage{iopams}

\def\be{\begin{equation}}
\def\ee{\end{equation}}
\def\ba{\begin{eqnarray}}
\def\ea{\end{eqnarray}}

\def\dd{\textrm{d}}

\def\Diff{\rm{Diff}}
\def\go{\mathring{g}}

\def\f{\frac}

\usepackage{graphicx}
\usepackage{graphicx,verbatim}
\usepackage{amsfonts}
\usepackage{amssymb}

\usepackage[colorinlistoftodos]{todonotes}
\usepackage{epstopdf}

\def\scri{\mathcal{I}}
\def\scrip{\scri^{+}}

\def\T{\mathcal{T}}
\def\scri{\mathcal{I}}

\newcommand{\pb}[1]{\hbox{\lower0.5ex\hbox{${}_{\leftarrow}$}}\kern-1.9ex{#1}}

\def\B{\mathfrak{B}}

\begin{document}

\title{Implications of a positive cosmological constant for general relativity}

\author{Abhay Ashtekar}
\address{Institute for Gravitation \& the Cosmos, and
    Physics Department, Penn State, University Park, PA 16802,
    USA, and\\
    CPT, Aix-Marseille Universit\'{e}, CNRS, F-13288 Marseille, France}
\ead{ashtekar@gravity.psu.edu}

\begin{abstract}

Most of the literature on general relativity over the last century assumes that the cosmological constant $\Lambda$ is zero. However, by now independent observations have led to a consensus that the dynamics of the universe is best described by Einstein's equations with a small but positive $\Lambda$. Interestingly, this requires a drastic revision of conceptual frameworks commonly used in general relativity, \emph{no matter how small $\Lambda$ is.}  We first explain why, and then summarize the current status of generalizations of these frameworks to include a positive $\Lambda$, focusing on gravitational waves.

\end{abstract}
%\submitto{\CQG}
%\maketitle

\section{Introduction}
\label{s1}

Hundred years ago, Einstein brought about a paradigm shift in our understanding of space, time and gravitation. Perhaps the most striking implications of the new paradigm occur via emergence of gravitational waves as ripples in spacetime geometry; through the formation of black holes due to gravitational collapse; and in cosmology where spacetime geometry now becomes dynamical. Literature in these areas has generally used Einstein's equations with zero cosmological constant, $\Lambda$. However, by now there is strong observational evidence that `dark energy' dominates the energy budget of the universe \cite{sndata,planck}. The dynamical effect of dark energy is an accelerated expansion. The simplest --and currently the best-- strategy is to model dark energy via a `small' but positive $\Lambda$. In this paper we adopt this viewpoint for brevity. However, our considerations would remain valid if there were another mechanism responsible for the accelerated expansion, so long as this expansion continues to the infinite future. 

It turns out that presence of a positive $\Lambda$ has a deep conceptual impact on all three areas mentioned above because the limit $\Lambda\to 0$ is discontinuous. New and qualitatively different structures appear if $\Lambda$ is positive, \emph{no matter how small it is,} requiring us to revise the very foundations of our understanding of several aspects of strong field gravity \cite{abklett}. The purpose of this Key Issues Review is to bring these features to forefront.

Let us begin with standard cosmology with spatially flat, homogeneous, isotropic spacetimes. If $\Lambda$=0, the past of the world line of an eternal cosmic observer would cover all of spacetime, as in Minkowski space. If $\Lambda >0$, on the other hand, there are cosmological horizons: the past contains only a spatially finite portion of spacetime. At the surface of last scattering, this portion is a ball of radius $\sim\,17.3$ Mpc, while at the onset of inflation (say with the Starobinsky potential) the ball has radius $\sim 5.3 \times 10^{-26}$cm \cite{ag3}, some 13 orders of magnitude \emph{smaller than the size of a proton.} A cosmic observer will not be able to receive any signals sent from outside these balls, no matter how long she waits. Therefore, unlike in the $\Lambda$=0 case, initial conditions in this tiny ball at the onset of inflation determine everything that a cosmic observer can ever hope to observe!

\begin{figure}[]
  \begin{center}
  %\vskip-0.4cm
    %$a)$\hspace{8cm}$b)$
    \includegraphics[width=1.4in,height=2.2in,angle=0]{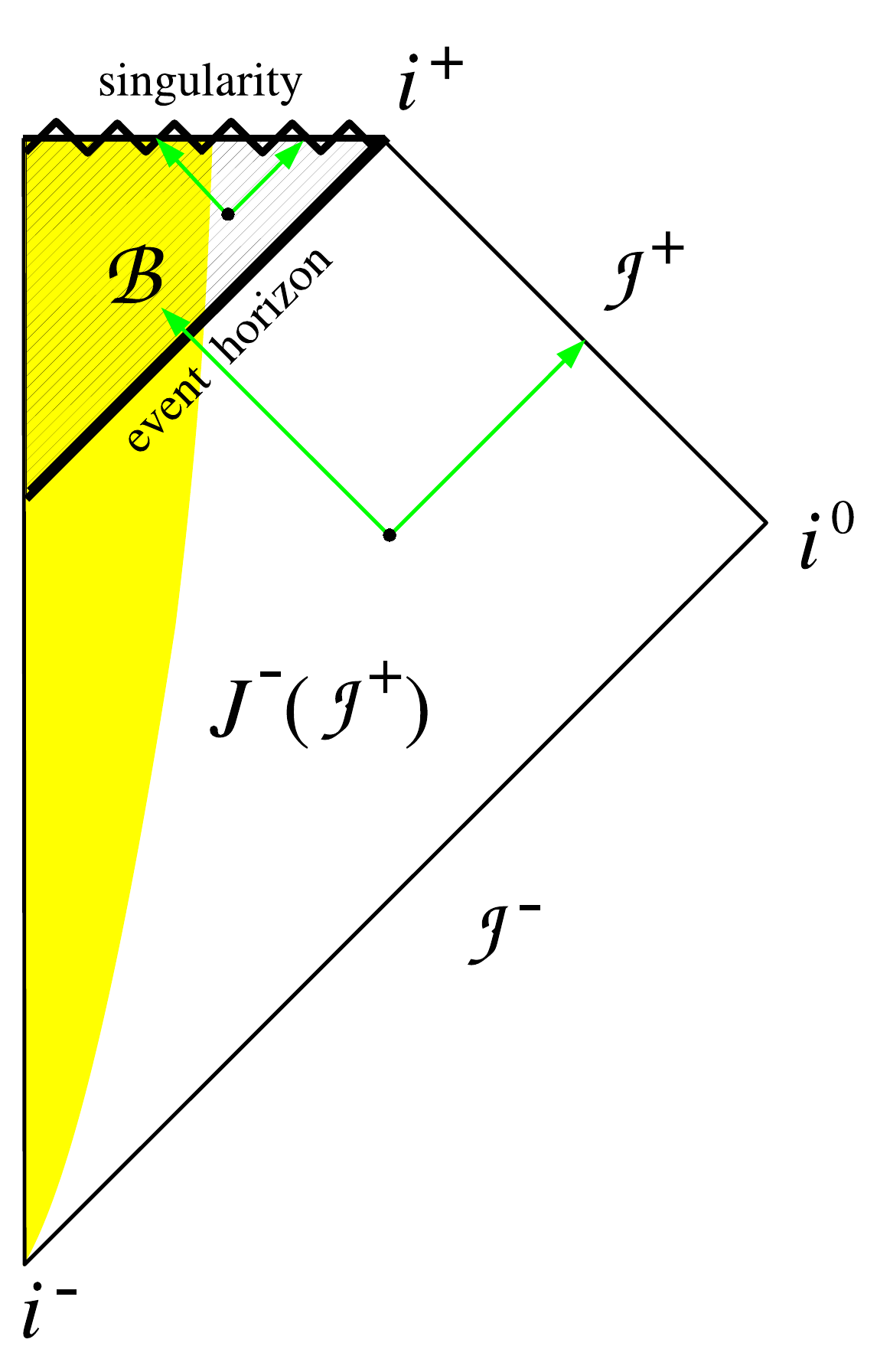}\hskip2.7cm
    \includegraphics[width=2in,height=2in,angle=0]{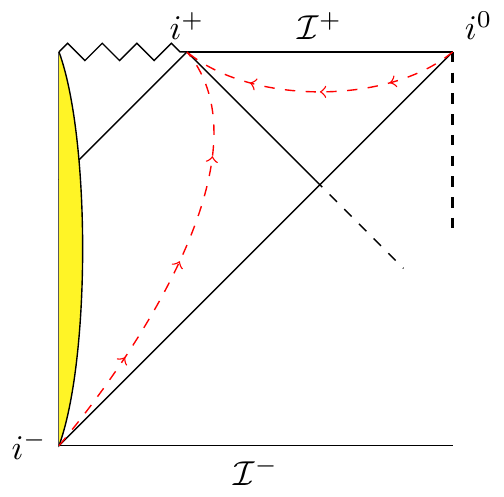}
\caption{Penrose diagrams depicting gravitational collapse of a spherical star \, {\bf A.}\, \textit{Left Panel:} The familiar $\Lambda$=0 case. $\mathcal{B}$ is the black hole region and (green) arrows show light rays  \, (credit: Jose Jaramillo and Eric Gourgoulhon). {\bf B.}\,\textit{Right Panel:} $\Lambda>0$ case. The collapsing body is visible only from points in the future of the cosmological horizon represented by the diagonal line. Dashed (red) arrows represent integral curves of the static Killing field.}
\label{collapse}
\end{center}
\end{figure}

For black holes, one also encounters unforeseen situations \cite{num2,abk1}. Fig. 1 shows the difference in the $\Lambda$=0 case (left panel) and $\Lambda >0$ case (right panel). In the second case, spacetime is incomplete to the right. This is because while the Kruskal extension of the Schwarzschild spacetime has a single black hole (and a single white hole) the complete Schwarzschild-de Sitter spacetime has an infinite number of black holes (and white holes). In the Schwarzschild-de Sitter case, the standard strategy is to arrive at a single black (and white) hole by using a discrete isometry to make an identification. This isometry is not available in the case of a single collapsing star. Since spacetime is `open' at the dashed line to the right, it is not sufficient to impose the  `no incoming radiation' condition on $\scri^{-}$ in the classical theory. Similarly in the analysis of the Hawking effect, one can no longer use $\scri^{-}$ to specify the incoming vacuum state.

Thus, unforeseen issues arise in cosmology and black hole physics. In this brief review we focus on gravitational waves whose discovery in 2015 by the LIGO collaboration has revolutionized the field. In section \ref{s2} we summarize the key \emph{conceptual} difficulties introduced by a positive $\Lambda$ and in section \ref{s3} the current status of their resolution. Section \ref{s4} illustrates the outlook through examples. 

\section{Gravitational waves: Even a tiny $\Lambda$ casts a long shadow}
\label{s2}

Let us begin by recalling how gravitational waves are described in the $\Lambda$=0 case. Already in the years 1916-18, Einstein showed that general relativity (GR) admits gravitational waves \emph{in the linearized approximation} and derived the celebrated quadrupole formula. However, two decades later, he suggested that this result was an artifact of linearization and gravitational waves do not exist in the full, nonlinear theory \cite{Kennefick1}! Confusion on the reality  of gravitational waves in full GR persisted until the 1960s \cite{Kennefick2} primarily because what seemed like `wave-like propagation' in one coordinate system could disappear in another. It was finally dispelled through the work of Bondi, Sachs, Trautmann and others \cite{bondi-sachs,at} who introduced a conceptual framework to extract gauge invariant information in waves by moving away from sources in \emph{null} directions. Penrose \cite{rp1} geometrized this framework by introducing a conformal completion of spacetime with boundary, $\scri$, that represents null infinity and serves as the natural arena to analyze gravitational radiation. In particular, there is a coordinate invariant field on $\scri$, now called the \emph{Bondi news tensor} $N_{ab}$ \cite{aa-radmodes}, that characterizes the presence of gravitational waves. Thus, for example, the condition $N_{ab}=0$ on past null infinity $\scri^{-}$ succinctly captures the physical requirement that there is no incoming gravitational radiation. 
In addition, $\scri$ has become an essential ingredient in the description of isolated systems, particularly black holes, both in classical and quantum gravity.

The new framework also brought out an unforeseen feature. Spacetimes admitting Penrose's completion are asymptotically flat in the sense that the physical metric ${g}_{ab}$ approaches a Minkowski metric 
$\eta_{ab}$ in a precise manner. %\,\, $\dd \tilde{s}^{2}\to -\dd u^{2} - 2\dd{u} \dd{r} + r^{2}\,(\dd\theta^{2} + \, \sin^{2}\theta\,\dd \varphi^{2})$, as we recede from sources along null directions, 
However, in presence of gravitational waves ---i.e., when $N_{ab}\not= 0$ at $\scri$--- \, $\eta_{ab}$ is not unique. Given one such $\eta_{ab}$, we can obtain a new Minkowski metric $\eta^{\prime}_{ab}$ by performing an `\emph{angle dependent} translation', e.g. $t \to t^{\prime}=t + f(\theta,\phi)$, to which $g_{ab}$ asymptotes in the same manner. 
%Then $\t{g}_{ab}$ also approaches $\eta^{\prime}_{ab}$ in the same manner. in the obvious way. Then $\t{g}_{ab}$ also approaches $\eta^{\prime}_{ab}$ as $1/r$, satisfying again, all the fall-off conditions introduced by Bondi et al. 
As a result, the asymptotic symmetry group is not the Poincar\'e group but an infinite dimensional generalization $\B$ thereof, obtained, so to say, by consistently putting together Poincar\'e groups of all Minkowski metrics to which ${g}_{ab}$ approaches. $\B$ is called the Bondi, Metzner, Sachs (BMS) group. %It is again semi-direct product of the Lorentz group $\L$ with an Abelian group $\S$, but $\S$ is the group of all angle-dependent translations, called \emph{supertranslations}: $\B = \S \rtimes \mathcal{L}$. 
However, $\B$ does admit a unique, 4-dimensional Abelian normal subgroup $\T$, the group of translations \cite{sachs}, just as the Poincar\'e group does. Therefore the notion of energy-momentum is well defined at null infinity. 

\begin{figure}[]
  \begin{center}
  %\vskip-0.4cm
    %$a)$\hspace{8cm}$b)$
    \hskip0.3cm
    \includegraphics[width=1.2in,height=2.2in,angle=0]
    {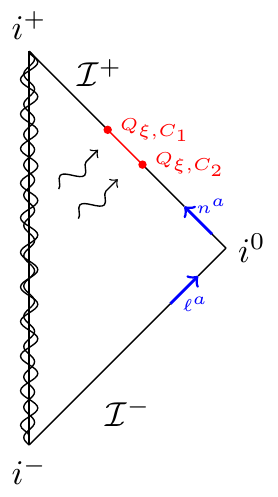} \hskip3cm
    \includegraphics[width=1.8in,height=1.8in,angle=0]{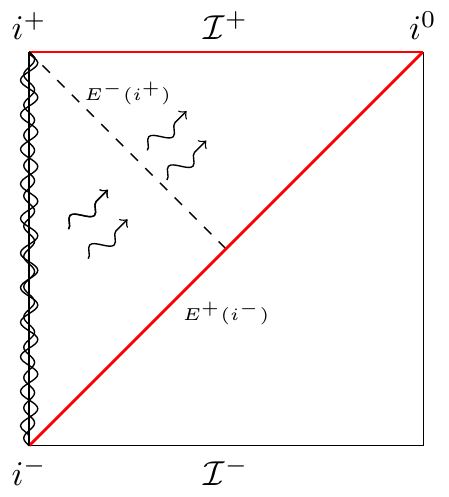}
\caption{Penrose diagrams of a binary system emitting gravitational waves.\, {\bf A.}\, \emph{Left Panel:} $\Lambda$=0 case, where $\scri$ is null.\, {\bf B.}\,\emph{Right Panel} $\Lambda >0$ case, where $\scri$ is space-like. $E^{+}(i^{-})$ is the past horizon and $E^{-}(i^{+})$ the future horizon for the star.}
\label{radiation}
\end{center}
\end{figure}

For definiteness, let us work with future null infinity, $\scrip$. A 2-sphere cross section $C$ of $\scrip$ (given by $u={\rm const}$ in Bondi coordinates) represents a `retarded instant of time'. Given a cross section $C$ and a BMS time-translation $t^{a}$ on $\scrip$, one can define a `gravitational charge integral' $Q_{t}[C]$ that represents the energy defined by $t^{a}$ at the retarded instant of time corresponding to $C$, called the Bondi-energy. There is a balance law: Given any two cross-sections $C_{1}, C_{2}$ on $\scrip$ as depicted on the left Panel of Fig. 2, the difference in the corresponding Bondi energies is given by \cite{bondi-sachs,rp1,aa-radmodes}:
\be \label{balance}\hskip-1.5cm Q_{t}[C_{2}] - Q_{t}[C_{1}] = \f{1}{\kappa}\,\int_{\Delta\scrip} \alpha |N_{ab}|^{2} \, \dd^{3}\scrip\, + \hbox{\rm matter energy flux} =: F_{t}[\Delta\scrip]\, , \ee
where $\Delta\scrip$ is the region of $\scrip$ bounded by $C_{1}$ and $C_{2}$ and $\alpha >0$ is a function representing the given BMS translation $t^{a}$. The right hand side, the flux $F_{t}[\Delta\scrip]$ of the Bondi energy carried by gravitational waves across $\Delta\scrip$, is manifestly positive. %Finally a more recent important result in GR is that in any regular, asymptotically flat solution of Einstein's equations, with matter satisfying the dominant energy condition, the Bondi energy $Q_{t}[C]$ is is positive, vanishing if and only if spacetime is flat \cite{hp-sy}. 
These results established reality of gravitational waves in full GR and provided powerful tools to extract physics from dynamics of isolated systems. For example, in simulations of black hole mergers, one calculates the Bondi 3-momentum carried by gravitational waves to determine the `kick' received by the final black hole \cite{kicks}.

%However while the Poincar\'e group admits only a 4-parameter family of Lorentz subgroups (related by translations), $\B$ admits an infinite parameter family (related by supertranslations). Consequently, the notion of angular momentum now acquires a `supertranslation ambiguity'.

The Bondi-Sachs framework was introduced half a century ago \cite{bondi-sachs}. Yet, it has still not been fully extended to the $\Lambda >0$ case because the rich structure at $\scrip$ used in the $\Lambda$=0 theory does not admit a direct generalization. Some of the conceptual difficulties have been noted over the last decade (see, e.g., \cite{bicak,rp2}). However, the full extent of the problem became clear only two years ago \cite{abk1} when systematic attempts at extending the framework to the $\Lambda >0$ case began. The extension of Einstein's quadrupole formula was obtained only last year and it is only very recently that an analog of the Bondi news was identified using cosmological horizons $H^{\pm}$ as `local $\scri^{\pm}$ (See Fig. 3.) It provides a gauge invariant characterization of gravitational waves at $H^{+}$. %But, for positive $\Lambda$, we do not yet have a similar characterization at $\scri^{\pm}$.

The first key difference is that while $\scrip$ is null if $\Lambda=0$, it is space-like if $\Lambda >0$ irrespective of its value (see the right panel in Fig. 2). Thus the limit $\Lambda \to 0$ is discontinuous. This fact has deep implications for asymptotic symmetries. Since the normal $n^{a}$ to a null surface is also tangential to it, in the $\Lambda$=0 case $\scrip$ comes naturally ruled by the integral curves of $n^{a}$. Therefore, asymptotic symmetries have to preserve this ruling. Secondly, because $\scrip$ is null, the intrinsic metric $q_{ab}$ on $\scrip$ is in effect a 2-sphere metric and therefore belongs to the unique conformal class of metrics that a 2-sphere admits. This additional structure reduces the asymptotic symmetry group from $\Diff (\scrip)$, the group of \emph{all} diffeomorphisms of $\scrip$, to $\B$, which has richer structure. By contrast, in the $\Lambda >0$ case, since $\scrip$ is space-like, its normal is transverse to it. So we lose the ruling; the intrinsic metric $q_{ab}$ now has signature (+,+,+), whence its conformal class is arbitrary. Therefore, the asymptotic symmetry group is now full $\Diff(\scrip)$. In particular, it does not admit a preferred 4-dimensional subgroup that can be thought of as `translations'. Consequently, we lose the ability to unambiguously identify energy-momentum: We have neither the analog of the Bondi 4-momentum `charges' nor the analog of the energy-momentum carried away by gravitational waves. 

One might imagine generalizing the construction and associating `charge integrals' $Q_{\xi}[C]$ with \emph{any} vector field $\xi^{a}$ on $\scrip$, i.e., with every generator of $\Diff(\scrip)$. Furthermore, field equations naturally suggest a candidate \cite{strominger,abk1}:
\be \label{charge} Q_{\xi}[C] = \Big(\f{3}{\Lambda}\Big)^{\f{1}{2}}\f{1}{8\pi G}\, \oint_{C} \big(\underline{E}_{ab} + \underline{T} q_{ab}\big) \, \xi^{a}\, \dd S^{b}\, , \ee
where $\underline{E}_{ab}$ and $\underline{T}$ are fields representing the asymptotic electric part of the Weyl tensor and trace of the matter stress-energy tensor. However, interpretation of these `charges' is obscure since $\xi^{a}$ is arbitrary. Furthermore, one finds that in the Schwarzschild-de Sitter spacetime which has neither matter nor gravitational waves, the `charges' $Q_{\xi}[C]$ are not conserved for a general $\xi^{a}$. This casts a serious doubt on their physical relevance.

Boundary conditions used in the $\Lambda>0$ case parallel those in the $\Lambda$=0 case \cite{rp1,abk1}. Nonetheless, while in the asymptotically flat case the physical metric $g_{ab}$ approaches a Minkowski metric --albeit not a unique one-- in the $\Lambda>0$ case, it \emph{need not} approach any de Sitter metric $\go_{ab}$ near $\scri$ \cite{hc,abklett}. Are the boundary conditions perhaps too weak? %This suggests that the boundary conditions may be too weak. 
A natural strategy is to strengthen them by asking, in addition, that the \emph{intrinsic metric $q_{ab}$ on} $\scrip$ be conformally flat, as in de Sitter spacetime. Then $g_{ab}$ does approach a de Sitter metric $\go_{ab}$ near $\scrip$. Furthermore, the asymptotic symmetry group is reduced from the infinite dimensional $\Diff(\scrip)$ to the 10 dimensional de Sitter group, $G_{\rm dS}$, allowing us to introduce the notion of de Sitter energy-momentum and angular momentum. In Kerr-de Sitter spacetimes only two of the ten `charges' of Eq. (\ref{charge}) are non-zero and they yield the expected `mass' and `angular momentum' \cite{abk1}. So, at first sight the strategy seems to be successful. However, now infinitesimal generators $\xi^{a}$ of $G_{\rm dS}$ are \emph{conformal Killing fields} of $q_{ab}$. As a consequence, in \emph{general} spacetimes which have no matter flux at $\scrip$, all ten de Sitter `charges' $Q_{\xi}[C]$ are \emph{absolutely} conserved, i.e., are independent of the choice of the cross-section. In this class of space-times, gravitational waves in them cannot carry away energy, momentum or angular momentum!

To summarize, if we do not strengthen boundary conditions we have no way of identifying quantities such as energy-momentum `charges' and fluxes, needed to extract physics of the given isolated system. Alternatively, we can strengthen the boundary conditions and speak of de Sitter energy-momentum and angular momentum. But now these quantities cannot be radiated away, signaling that the restriction is unreasonably severe. There is a precise sense in which they are the $\Lambda >0$ analogs of asymptotically flat spacetimes in which the Bondi news $N_{ab}$ vanishes identically at $\scrip$ \cite{abk1}. (In particular, in this sub-class of asymptotically flat spacetimes, the BMS group $\B$ reduces to the Poincar\'e group, just as in the $\Lambda>0$ case $\Diff(\scrip)$  reduces to the de Sitter group \cite{aa-radmodes}.) Thus, there is a quandary if $\Lambda$ is positive, no matter how tiny.

\section{Gravitational waves with $\Lambda >0$: Current status}
\label{s3}

In this section we will sketch the current status of the subject through a few illustrative results, first in the linearized approximation and then in full, non-linear GR. 
\subsection{Linearized theory}
\label{s3.1}
1. \emph{Difficulties:} We now have to linearize GR off de Sitter metric $\go_{ab}$ rather than Minkowski. Can we require that linearized perturbations should preserve conformal flatness of de Sitter $\scrip$ to first order? Unfortunately this condition removes, by hand, half the linearized fields which, in the language used in the cosmological perturbation theory, correspond to `growing modes'. Furthermore, one finds that the remaining perturbations do not carry fluxes $F_{\xi}$ across $\scrip$ for \emph{any} generator $\xi^{a}$ of $\Diff(\scrip)$ \cite{abk2}. Thus, the requirement is too severe already in the linear approximation. However, even without this requirement, we now have a preferred subgroup $G_{\rm dS}$ of $\Diff (\scrip)$, induced by the isometry group of $\go_{ab}$. Thus, the quandary we encountered in the full theory is now bypassed because we have a de Sitter background. 

But since $\scrip$ of $\go_{ab}$ is space-like, other difficulties persist \cite{abk2,abk3}. For example: \\
\noindent (i) Every Killing field of $\go_{ab}$ is space-like in a neighborhood of $\scrip$ including the one which represents a time translation near the source (see Fig. 1B). Therefore, in stark contrast with the situation in the $\Lambda=0$ case, linearized gravitational (\emph{as well as electromagnetic}) waves can carry unboundedly large \emph{negative} energy across $\scrip$.\\
(ii) In deriving the $\Lambda=0$  quadrupole formula, one makes heavy use of $1/r$ expansions and calculates the energy flux across $r$=const\, time-like cylinders which asymptote to $\scrip$ in the large $r$ limit. %In particular, the `transverse-traceless decomposition' used  in this calculation is gauge invariant only up to $\mathcal{O}(1/r^{2})$ terms (see, e.g. \cite{pw}). 
In de Sitter spacetime, by contrast, such time-like cylinders approach the past cosmological horizon $E^{+}(i^{-})$ rather than $\scrip$. For retarded solutions of interest, the energy flux across $E^{+}(i^{-})$ vanishes identically. Thus, a new approximation scheme tailored to the de Sitter $\scrip$ is needed.\\
(iii) Because of the expansion of the universe, physical wavelengths of gravitational (and electromagnetic) waves grow as they propagate away from sources and can vastly exceed the curvature radius in the asymptotic region, making the standard high frequency expansions (and the geometric optics approximation) untenable near $\scrip$. \smallskip
 
2. \emph{The quadrupole formula:} Because of such unforeseen difficulties, the problem of generalizing Einstein's quadrupole formula had remained open for a century. It has now been resolved by appropriately modifying Einstein's calculation to address these issues \cite{abklett,abk3}. Specifically, one has to replace the $1/r$ approximation with a suitable `late-time', post-de Sitter approximation, and restrict oneself to sources that are `isolated' in the sense that they remain within a spatially bounded world-tube whose physical radius is smaller than the cosmological radius. Also, the derivation does \emph{not} make use of a high frequency approximation. The end result is that Einstein's formula $P_{t}(u_{0}) = \frac{G}{8 \pi} \oint_{u=u_{0}}\!\! {\rm d}^{2}S\, |{\buildrel {...} \over{Q}}_{ab}^{({\rm TT})}{(\rho)}|^{2}$,\, for power emitted at a retarded time $u=u_{0}$ is replaced by \cite{abk3}

\be \label{power}
P_t (u_{0})\,\, =\,\, \frac{G}{8 \pi} \oint_{u=u_{0}}\!\! {\rm d}^{2}S\,\Big[\mathcal{R}^{ab}\, \mathcal{R}_{ab}^{(\rm{TT})}\,\Big]\, ,\ee
where the `radiation field' $\mathcal{R}_{ab}$ on $\scrip$ is given by\\ 
\centerline{$\mathcal{R}_{ab}\, =\, \Big[{\buildrel {...} \over{Q}}_{ab}^{(\rho)} + (3\Lambda)^{\f{1}{2}}\, \ddot{Q}_{ab}^{(\rho)} + (2\Lambda/3)\, \dot{Q}^{(\rho)}_{ab} + (\Lambda/3)^{\f{1}{2}}\,  \ddot{Q}_{ab}^{(p)} + \Lambda\, \dot{Q}_{ab}^{(p)} + 2 (\Lambda/3)^{\f{3}{2}}\, Q_{ab}^{(p)}\Big](u_{0}). $}
\noindent Here ${\rm (TT)}$ stands for `transverse-traceless'. The `pressure quadrupole moment' ${Q}_{ab}^{(p)}$ is obtained by substituting pressure in place of density in the standard `mass' quadrupole moment ${Q}_{ab}^{(\rho)}$. As in the $\Lambda=0$ case, the integral is over the 2-sphere cross-section of $\scri^{+}$ defined by the retarded time $u=u_{0}$, and the center of mass of the source follows an integral curve of the world line of the Killing field $t^{a}$ used to define energy and power.

3. \emph{New features:} This analysis brings out the following interesting points: (i) From cosmology we know that unlike in Newton's theory, pressure gravitates in general relativity; now we learn that it also sources gravitational waves already at the lowest post-Newtonian order. \, (ii) Because energy is associated with a Killing field $t^{a}$ of $\go_{ab}$, and $t^{a}$ is (null and) future-directed on the cosmological horizon $E^{-}(i^{+})$, the energy flux across  $E^{-}(i^{+})$ is positive (see Fig. 2B). Because we are considering retarded solutions, there is no flux across the past horizon of the source, $E^{+}(i^{-})$. Finally, flux of the $t$-energy across $\scrip$ equals that across $E^{-}(i^{+})$ because energy associated with a Killing field is conserved. Hence power radiated across $\scrip$ by a physical source is necessarily positive even though in general gravitational waves can carry negative energy.\, (iii) Eq. (\ref{power})  provides $\Lambda$-corrections to Einstein's formula. In particular, it tells us from first principles that if the dynamical time scale $\tau$ associated with the source is small compared to $1/\sqrt{\Lambda}$, the error involved in neglecting the presence of $\Lambda$ is small, of order $\mathcal{O}(\tau \sqrt{\Lambda})$, even though the  limit $\Lambda \to 0$ is fundamentally discontinuous.

Thus, in spite of the fact that the $\Lambda>0$ framework is conceptually very different from the more familiar one with $\Lambda=0$, general expectations based on physical intuition are borne out, but now from first principles, with a \emph{quantitative} control on errors one makes by setting $\Lambda=0$. 

\subsection{Full general relativity}
\label{s3.2}
1. \emph{No incoming radiation condition:} We cannot mimic the strategy of incorporating this condition by requiring $N_{ab}=0$ at $\scri^{-}$ because for $\Lambda>0$ we do not have the analog of Bondi news, $N_{ab}$ there. But since isolated systems of interest remain in a spatially bounded world-tube, they pierce $\mathcal{I}^{\pm}$ at single points, $i^{\pm}$. It is clear e.g. from Fig 2B that an observer in the triangular region below the cosmological horizon $E^{+}(i^{-})$ cannot receive a causal signal from the source. Therefore, to study this isolated system, it suffices to restrict oneself to the upper triangle and ask that there be no incoming radiation at its past boundary, $E^{+}(i^{-})$. Fortunately, since $E^{+}(i^{-})$ is null, we can make use of the `isolated horizons' framework \cite{akrev} to do so. 

Detailed investigation shows that the natural way to impose the requirement that there be no incoming gravitational waves (or matter flux) is to ask that  $E^{+}(i^{-})$ be a \emph{weakly isolated horizon} (WIH) \cite{abl1}. %(These considerations are subtle; requiring $E^{+}(i^{-})$ to be an isolated horizon would be too strong!) 
Thus, in the $\Lambda>0$ case, we can entirely forego $\scri^{-}$. Note that $E^{+}(i^{-})$ is a WIH not only in Kerr-de Sitter spacetimes but also in numerical studies including the stellar collapse \cite{num2} depicted in Fig. 1B. Therefore, this strategy also neatly bypasses the difficulty illustrated in Fig. 1B: If $E^{+}(i^{-})$ is a WIH, we are guaranteed that the outgoing radiation near $\scrip$ is not contaminated by anything entering from the region in the past of the cosmological horizon $E^{+}(i^{-})$.
%across the unspecified dashed line: incompleteness of spacetime to the right of the dashed line becomes irrelevant. 

2. \emph{Symmetries and `charges' at the past boundary:} The strategy of  using $E^{+}(i^{-})$ as the past boundary in place of $\scri^{-}$ also resolves this issue. The WIH structure enables one to single out a 
time translation symmetry $t^{a}$, and define the associated energy $Q_{\xi}[C]$ which is independent of the choice of the cross-section $C$ 
because there is no matter or gravitational radiation flux across $E^{+}(i^{-})$ \cite{akrev}. If the intrinsic geometry of $E^{+}(i^{-})$ is axi-symmetric, one can also define mass and angular multipoles \cite{aepv} that carry  detailed information about the source configuration in the distant past.
\begin{figure}[]
  \begin{center}
  %\vskip-0.4cm
    %$a)$\hspace{8cm}$b)$
    \includegraphics[width=4in,height=1.7in,angle=0]
    {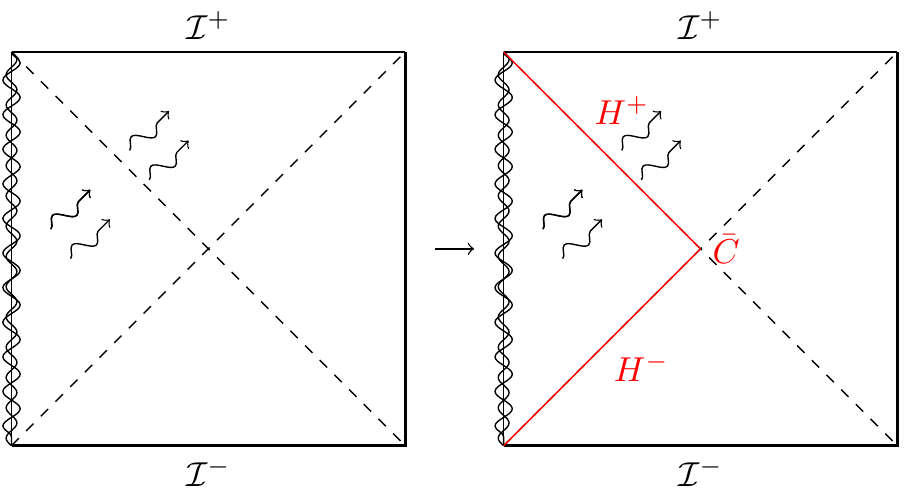} 
\caption{Proposal: Use $H^{\pm}$ as the arena to study gravitational waves in place of $\scri^{\pm}$.}
\label{proposal}
\end{center}
\end{figure}

3. \emph{`Local' $\scri^{\pm}$}: In the discussion of the quadrupole formula, the energy flux could be evaluated at $\scri^{+}$ because we could isolate the appropriate `time translation' $t^{a}$ using the background de Sitter metric $\go_{ab}$. In full GR this has not been possible because, as we saw in section \ref{s2}, the physical metric need not approach $\go_{ab}$ even at $\scri^{+}$. Recall, however, that in the linear approximation the energy flux could also be evaluated on $H^{+}$, the future half of the cosmological horizon $E^{-}(i^{+})$. This motivates the idea of replacing $\scri^{\pm}$ with $H^{\pm}$ as the arena for gravitational waves in the $\Lambda >0$ case (see Fig. 3). We can regard $H^{\pm}$ as the `local $\scri^{\pm}$, tailored to the isolated source under investigation.

With this replacement, the spacetime region of interest --the left triangle of Fig. 3-- is very similar to the asymptotically flat spacetime, depicted in Fig. 1A. Also, unlike $\scri^{\pm}$ for $\Lambda>0$, $H^{\pm}$ are null and ruled by their null normals, just as $\scri^{\pm}$ are for $\Lambda=0$. But there are also key differences because $H^{\pm}$ are proper submanifolds of the physical spacetime, rather than boundaries. As a consequence, the structure on $H^{-}$ is more rigid than that on $\scri^{-}$ in the $\Lambda=0$ case. In particular, it carries a preferred time translation.
%In the $\Lambda=0$ case, the symmetry group at $\scri^{-}$ reduces to the Poincar\'e group because $N_{ab}=0$ there. For $\Lambda>0$, because of the WIH conditions\, --and the presence of a preferred cross-section $\bar{C}$ of $H^{-}$ at which it intersects $H^{+}$--\, the symmetry group can be reduced to 4-dimensional sub-group $\G$ of $G_{\rm ds}$.
The structure on $H^{+}$, on the other hand, is \emph{less rigid} than that on $\scri^{+}$ in the $\Lambda=0$ case: while the intrinsic metric on $\scri^{+}$ is Lie dragged by the null normal $n^{a}$, the intrinsic metric on $H^{+}$ is truly dynamic. %Therefore, if one were to use only the physical geometry of $H^{+}$, one would be left with a symmetry group that is larger than the BMS group $\B$.  However,  
Nonetheless, since $H^{+}$ intersects $H^{-}$ in a 2-sphere cross-section $\bar{C}$, one can systematically `drag' the the time translation from $H^{-}$ to $H^{+}$. This is is a delicate, well-defined procedure, guided by physical requirements.                                                                                                                                                                                                                                                                                                                                                                                                                                                                                                                                                                                                                                                                                                                  

With this structure at hand, one uses an action principle based on null boundaries \cite{ww} to define energy $Q_{t}[C]$ associated with 2-sphere cross sections of $C$:  \vskip0.25cm
\centerline {$\hskip4cm Q_{t}[C] =\, \f{1}{\kappa}\, \oint_{C} \big[\Theta_{t} + \kappa_{t}\big]\, ,$\hskip6.2cm (4)} 

\noindent where $\Theta_{t}$ and $\kappa_{t}$ are the expansion and the surface gravity defined by the `time-translation symmetry' $t^{a}$ on the horizon $H^{+}$,\, and the balance law\vskip0.2cm
\centerline{$ Q_{t}[C_{2}] - Q_{t}[C_{1}] = \f{1}{\kappa}\, \int_{\Delta{H}^{+}} \alpha\, \big{|}\sigma^{(n)}_{ab} \big{|}^{2} \, \dd^{3}H^{+}\,\,+\, \hbox{\rm matter energy flux}\,\,=:\, F_{t} [\Delta H^{+}]\, .\qquad (5)$}
\vskip0.2cm
\noindent Here $t^{a} = \alpha n^{a}$ on $H^{+}$, where $n^{a}$ is the geodesic vector field tangential to $H^{+}$, normalized using the structure induced on $\bar{C}$ by $H^{-}$. Note that the integrand in the flux expression is positive definite. Comparison with (\ref{balance}) tells us that $\sigma^{(n)}_{ab}$ is the analog of the Bondi news $N_{ab}$ in the $\Lambda=0$ case: \emph{the condition  $\sigma_{ab}^{(n)}|_{H^{+}}\not= 0$ now provides us the desired gauge invariant characterization of gravitational waves at $H^{+}$} --the local $\scri^{+}$. In a carefully taken $\Lambda \to 0$ limit, these expressions reproduce the standard `charge integral' and balance law for Bondi energy at $\scri^{+}$ in the $\Lambda=0$ case. Also, the `charge' and flux integrals (4) and (5) bear out physical expectations in the Vaidya evaporation of a white hole, even though $H^{+}$ is dynamical, with teleological features because of the matter flux. 

Results reported in Sec. \ref{s3.2}, as well as those on a Hamiltonian formulation of GR at $\scri^{+}$ of asymptotically de Sitter spacetimes are being written up for publication. As of now there is no completely satisfactory characterization of gravitational waves at $\scri^{+}$.

\section{Discussion}
\label{s4}

A key feature of GR with $\Lambda >0$ is that no matter how far one recedes from an isolated body, in contrast to the $\Lambda=0$ case, spacetime curvature does not go to zero. Consequently, much of our well-developed intuition from asymptotically flat space-times does not carry over. Another difference lies in the topology of $\scri^{\pm}$. In the $\Lambda=0$ case, it is $\mathbb{S}^{2}\times\mathbb{R}$ for any isolated system, just as in Minkowski space. With $\Lambda>0$, the topology is $\mathbb{S}^{3}$ in de Sitter spacetime, %$\mathbb{R}^{3}$ for Friedmann-Lema\^itre spacetimes, 
but $\mathbb{S}^{2}\times \mathbb{R}$ for an isolated star or black hole \cite{abk1}. Because of such qualitative differences, one cannot directly use the powerful mathematical results on non-linear stability of de Sitter space-times \cite{hf} in the analysis of physical properties of isolated systems with $\Lambda>0$. 

Such conceptual differences also give rise to new features in the theory of black holes. The issue of uniqueness of Kerr-de Sitter black holes has still not been established in 4 spacetime dimensions. Furthermore, non-rotating black holes also acquire new features. First, because the space-time metric does not approach de Sitter metric near $\scrip$ in presence of radiation, unforeseen complications arise in the analysis of non-linear stability of the Schwarzschild de Sitter space-time \cite{schlue,hintz-vasy}. Also, these black holes have an unforeseen property: there is an upper bound on their mass: $M_{\rm max} = 1/3G\big(\Lambda)^{1/2}$. Numerical simulations show that attempts to form black holes of higher mass fail \cite{num1,num3}. In particular if tries to achieve this by colliding gravitational waves,
%, if one attempts to form a black hole of higher mass by sending in gravitational waves, 
they simply disperse even when they have large amplitudes. The limit is observationally viable since, for currently accepted value of $\Lambda$,\,\, $M_{\rm max}$ is 12 orders of magnitude larger than the mass of the heaviest supermassive black holes we know. As we noted in section \ref{s1}, unforeseen issues arise also in the discussion of black hole evaporation: One appears to be stuck in the very first step of Hawking's original analysis since $\scri^{-}$ is no longer the appropriate arena to specify the incoming vacuum state in the spacetime of a collapsing star.  However, one can specify the incoming vacuum using $H^{-}$  --i.e., the `local $\scri^{-}$' (see Fig. 3). Because there is a time translation group on $H^{-}$, there is a well-defined notion of positive and negative frequencies to define the incoming vacuum. It would be most interesting to see how the value of $\Lambda$ enters the final density matrix on $H^{+}$ or $\scrip$. 

Returning to gravitational waves, there is a number of open issues in mathematical and numerical GR, geometrical analysis, and approximation methods. We will conclude with an example. Is the energy `charge integral' (4) positive under suitable physical restrictions? %If so, the `total energy' defined on $H^{-}$ would be automatically positive. But it may be easier to establish positivity directly on $H^{-}$, using a spinorial argument a la Witten. 
These questions on positivity also arise at $\scrip$ where the `charge integrals' can be introduced using a Hamiltonian formulation of GR. There  \emph{are} positive energy theorems in the $\Lambda>0$ case in the literature. However, they typically refer to the absolutely conserved `charge' at spatial infinity where gravitational waves do not reach \cite{ad,km,cjk}. Furthermore, `energy' they refer to is associated with conformal --rather than a time translation-- symmetry. Therefore, while the notion can be useful in mathematical analysis, its physical meaning is unclear even in the Maxwell theory. This is \emph{not} the energy that is related to the properties of sources --time derivatives of dipole moments in the Maxwell theory and of quadrupole moments in GR. The generalization (\ref{power}) of Einstein's quadrupole formula refers to the energy associated with a time-translation symmetry.  Finally, in linearized GR we have explicit expressions of corrections to Einstein's quadrupole formula. In full GR, the theory of gravitational waves using $H^{\pm}$ as `local' $\scri^{\pm}$ is becoming mature. But since it requires that $H^{+}$ be `sufficiently long' to intersect $E^{+}(i^{-})$, 
we do not yet know if this approach allows a sufficiently large class of examples. The Hamiltonian framework based on $\scrip$, by contrast, is free of this potential limitation but so far it has not enabled one to obtain expressions of local fluxes of energy carried by gravitational waves. As Eq. (\ref{power}) suggests, corrections to the $\Lambda=0$ theory are likely to be negligible for sources of interest to the current gravitational wave detectors. But it is possible that subtle effects induced by $\Lambda$ could be measured in the future \cite{abklett}. From a theoretical perspective, there is the more compelling motivation to address these issues: Since the accelerated expansion of our universe is now well established, at a fundamental level we need to know how to characterize gravitational waves and understand their properties \emph{within this paradigm,} before developing approximation methods, however important they may be in practice.
%\vfill\break

\textbf{Acknowledgments:} I would like to thank B\'eatrice Bonga and Aruna Kesavan for numerous discussions, for figures 2 and 3, and for their comments on the first draft of this manuscript; Wolfgang Wieland for discussions on his action principle with null boundaries; and Jose Jaramillo and Eric Gourgoulhon for permission to use figure 1.A.
This work was supported in part by the NSF grant PHY-1505411 and the Eberly endowment.

\end{document}